\documentclass[aps,twocolumn,groupedaddress,showpacs]{revtex4}
\usepackage{graphics}

\begin{document}
\bibliographystyle{apsrev}

\title{\bf Measuring Rashba spin orbit interaction strength without 
magnetic field }

\author{T.~P.~Pareek}
\affiliation{
Harisch Chandra Research Institute, Chaatnag Road, Jhunsi, Allahabad - 211019, India}

\begin{abstract}
We propose a multi-terminal(Non-magnetic) mesoscopic  electrical 
transport measurement for measuring Rashba spin orbit strength. 
The method proposed is different from the
usual magneto transport measurements, which require magnetic fields for e.g.
Shubnikov-de Hass oscillations or weak anti-localization measurements. 
Our methods uses three terminal mesoscopic device, 
in which one of the terminal acts as 
a voltage probe. We show that the Voltage measured at the non-magnetic 
voltage probe depends on the strength of Rashba spin-orbit interaction.


\end{abstract}
\maketitle

A central theme of spintronics research is to realize Datta-Das 
spin field effect transistor(Spin-FET) or its equivalent which would allow
manipulation of spin at mesoscopic level\cite{datta}. 
Datta-Das Spin-FET relies on the
Rashba spin-orbit interaction which can be controlled by applying
a gate voltage \cite{rashba}. Accordingly it is know as gate 
controlled spin-orbit interaction. The key idea is application of
gate voltage on the top of device
causes changes in asymmetry of confining potential and hence
controls Rashba-spin-orbit (RSO) interaction which is dependant on the interface electric field. This gate control of RSO interaction have been 
experimentally verified in many two dimensional 
narrow gap III-IV semiconductor heterostructures 
for e. g. in InAs/AlSb \cite{heida},
in GaSb-InAs-GaSb quantum wells  \cite{luo} and in In$_{x}$Ga{$_{1-x}$}As/
In$_{0.52}$Al$_{0.48}$As heterostructures by Nitta and Das {\it et al.}
\cite{nitta}. Experimentally,
Shubnikov-de Hass (SdH) oscillation are measured as a function of magnetic
field and applied gate voltage. 
Zero field spin splitting manifest itself as a beating pattern in 
SdH oscillation due to two close frequency  components with similar 
amplitudes arising from spin-split levels. The spin splitting at zero field
is extracted from finite field data by linear extrapolation \cite{lommer}.
However due to finite magnetic field, Zeeman splitting is always present
which makes it ambiguous to extract the spin orbit parameter. 
Further the top gate voltage also changes the concentration of charge carriers in spin sub bands. This affects the beating pattern in SdH oscillation which
crucially depends on the difference of carrier concentration between tow 
spin sub-bands. In-fact SdH oscillations vanishes for higher values of top gate voltage, since the carrier concentration in the channel saturate without filling
the second sub-band \cite{engels}. The spin-orbit coupling parameter thus can not be extracted beyond a certain gate voltage. Hence the measurement of Rashba
spin-orbit parameter with SdH oscillation is not free of ambiguities as was pointed out recently \cite{keppeler},\cite{tarasenko} \cite{averkiev}.

In light of above discussion, it is clear that it would be interesting and highly desirable if one can design an experiment which would allow to measure
the spin-orbit parameter with out the application of magnetic field. We 
propose and show in this letter that mesoscopic multi-terminal charge transport measurement provides such a way to measure sin-orbit interaction. To understand our proposal we recapitulate briefly the generalized Landauer-B\"uttiker 
theory for multi-terminal 
spin-charge transport 
developed by us recently \cite{pareek}.
For spin quantization axis to be along $\bf{\hat u}$, pointing along
($\theta$, $\phi$) where $\theta$ and
$\phi$ are usual spherical angles , the net spin and charge currents flowing
through terminal m are given by;

\begin{equation}
I_{m}^{q}=I_{m}^{\sigma} + I_{m}^{-\sigma} \\
\equiv \frac{e^2}{h}\sum_{n \neq m, \sigma, \alpha}\{(T_{n\, m}^{\alpha\,\sigma}V_{m}
-T_{m\, n}^{\sigma\, \alpha}V_{n}\}
\label{eq_ch}
\end{equation}
  
\begin{eqnarray}
I_{m}^{s}&=&I_{m}^{\sigma} - I_{m}^{-\sigma}  \nonumber \\
&\equiv& \frac{e^2}{h}\sum_{n \neq m,  \alpha}\{(T_{n\, m}^{\alpha\,\sigma}- T_{n\, m}^{\alpha\,-\sigma})V_{m}
+(T_{m n}^{-\sigma\, \alpha}- T_{m n}^{\sigma\, \alpha}) V_{n}\}
\label{eq_sc}
\end{eqnarray}

\noindent where $I_{m}^{q}$ is charge current and $I_{m}^{s}$ is spin
current flowing in terminal {\it m}. $T_{n\,m}^{\alpha, \sigma}$ are spin
resolved transmission coefficients from terminal {\it m} to {\it n} 
and $V_m$ is applied potential at terminal {\it m}. 

Now let us consider the special case where the voltages at terminal 1
and 2 are respectively 
$V_1$=0 and $V_{2}$ and terminal third
is a voltage probe, i.e, $I_{3}^{q}=0$ (see Fig.1, here we would like to stress that we consider a Y shaped three terminal conductor since it is symmetric with 120 deg rotation hence the effect due to geometrical asymmetry would be minimized which will not be the case if one considers other three terminal geometries for, e.g., a T shaped conductor). With this condition one
can determine the voltage, $V_{3}$, at third terminal using the set of
equation (\ref{eq_ch}) and is given by \cite{pareek1}, 
\begin{equation}
\frac{V_{3}}{V_{2}}=\frac{T_{32}}{T_{13}+T_{23}}
\label{volt}
\end{equation}

\noindent where $T_{n\,m}$ are total transmission coefficient
(summed over spin indices) from terminal {\it m} to {\it n}.

It is know from studies in mesoscopic physics that the presence of third lead can change the conductance, since it acts as a scatterer\cite{pareek1}. Hence in similar way if the scattering is changed within the sample that should also reflect in three terminal measurement.
This is seen clearly from (\ref{volt}) that the voltage $V_3$ at 
third non-magnetic
terminal depends on the ratio of transmission coefficients $T_{32}$ and
$T_{13}+T_{23}$. Presence of spin-orbit interaction does lead to spin conserved and spin flip scattering which in turn changes spin resolved transmission coefficient $T_{n\,m}^{\alpha, \sigma}$, and , hence the total transmission coefficient. Accordingly the voltage $V_{3}$ at third terminal will show changes as one
varies spin-orbit interaction strength through top gate voltage. 
{\bf In other words
$V_{3}$ will show dependence on the gate voltage.} So by calibrating these
changes one can extract the Rashba spin-orbit parameter. We would like to stress that we do not need to apply magnetic field which is the case for SdH
oscillation as well
for weak localization measurement. Hence this method will not have the
difficulties which was the case for other methods as discusses in introduction.

Here we would like to point out that In Ref.\cite{pareek} we had studied the
problem of how to measure the spin currents when terminal third is Ferromagnetic. Third terminal being Ferromagnetic
selectively allowed electrons whose
spins were parallel to the magnetization of lead, and, hence it was sensitive to the spin currents flowing. Further in ref. \cite{pareek} spin currents and corresponding voltage were studied as a function of quantization axis
for a fixed spin-orbit interaction strength. Since the origin of spin-orbit interaction was due to presence of impurity which is fixed for a given sample and can not be changed. 
While Rashba interaction can be changed through an external gate voltage, which is the topic of study in present manuscript.
Further in this article all leads are non-magnetic and there is 
no magnetic filed either. Since we are interested in the strength of spin-orbit interaction which depends on the ratio of total transmission coefficient in different leads.Hence in present study topic of investigation is charge transport when all the leads are non-magnetic  and how can one use it to measure strength of spin-orbit interaction
which is easier to measure experimentally as well is
different from spin current study as was done in Ref. \cite{pareek}.
Since leads are non-magnetic they are insensitive to different component of spin currents in other words even though the
spin currents flowing across {\it x}, {\it y} and {\it z} axis will be different but the voltage measured at third lead will be same and measures the strength of spin-orbit interaction. 

\begin{figure}
\resizebox{!}{2.5in}{\includegraphics{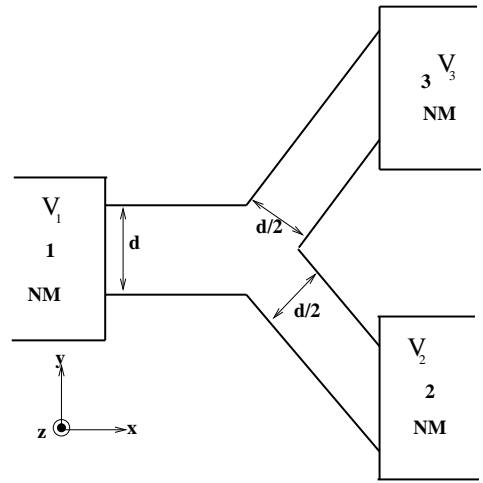}}
\caption{Y-shaped three terminal junction with applied voltages
$V_{1}$, $V_{2}$ and $V_{3}$ as depicted. Terminal third (labeled 3)
is a voltage probe and draws no charge current . 
All three terminals are non-magnetic.
}
\label{Fig. 1}
\end{figure}

To obtain quantitative results we perform numerical simulation on a
Y-shaped conductor shown in Fig. 1(we have discussed above that Y-shaped conductor being most symmetric among the three terminal devices which would minimized any spurious results due to geometrical asymmetry).
We model the conductor on a square tight binding lattice\cite{pareek2} 
with lattice spacing {\it a} and we use the corresponding tight
binding model including Rashba spin orbit interaction given by
\begin{equation}
H_{Rashba}=\alpha(\sigma_{x}k_{y}-\sigma_{y}k_{x})
\end{equation}
\noindent where $\sigma$'s are Pauli matrices and $k_{x}$ and $k_{y}$ are momentum vector along $x$ and $y$ axis respectively and $\alpha$ is Rashba spin-orbit interaction parameter.
For the calculation of spin resolved
transmission coefficient , we use the recursive green function
method. Details of this can be found in
Ref.\cite{pareek2}.
The numerical result presented takes the quantum effect
and multiple scattering into account and is exact at the single particle level.
For the model of disorder we take Anderson model,
where on-site energies are distributed randomly within [-U/2, U/2],
where U is the width of distribution.

To present quantitative results we measure energy
in units of hopping parameter
{\it t }for tight binding model, we set it equal to 1 above the band bottom.
Dimensionless Rashba coupling is defined as $\lambda_{rso}=\alpha/2*t*\lambda_{f}$, where
$\lambda_{f}$ is Fermi energy. Experimentally observed values of these parameters are  are 
$E_{f}\approx 80-120 meV$, corresponding $\lambda_{f}\approx 70-200 A^o$
and the RSO coupling $\alpha$ varies between $0.4\times 10^{-11} eV-m$ to
$1.2\times 10^{-11} eV-m$ \cite{heida,luo,nitta,das}. 
Accordingly our dimensionless Rashba coupling
parameter $\lambda_{rso}$ varies between 0.005 to 0.04. In our numerical simulation we have varied $\lambda_{rso}$ upto 0.05.

\vspace{1cm}
\begin{figure}
\resizebox{!}{2.in}{\includegraphics{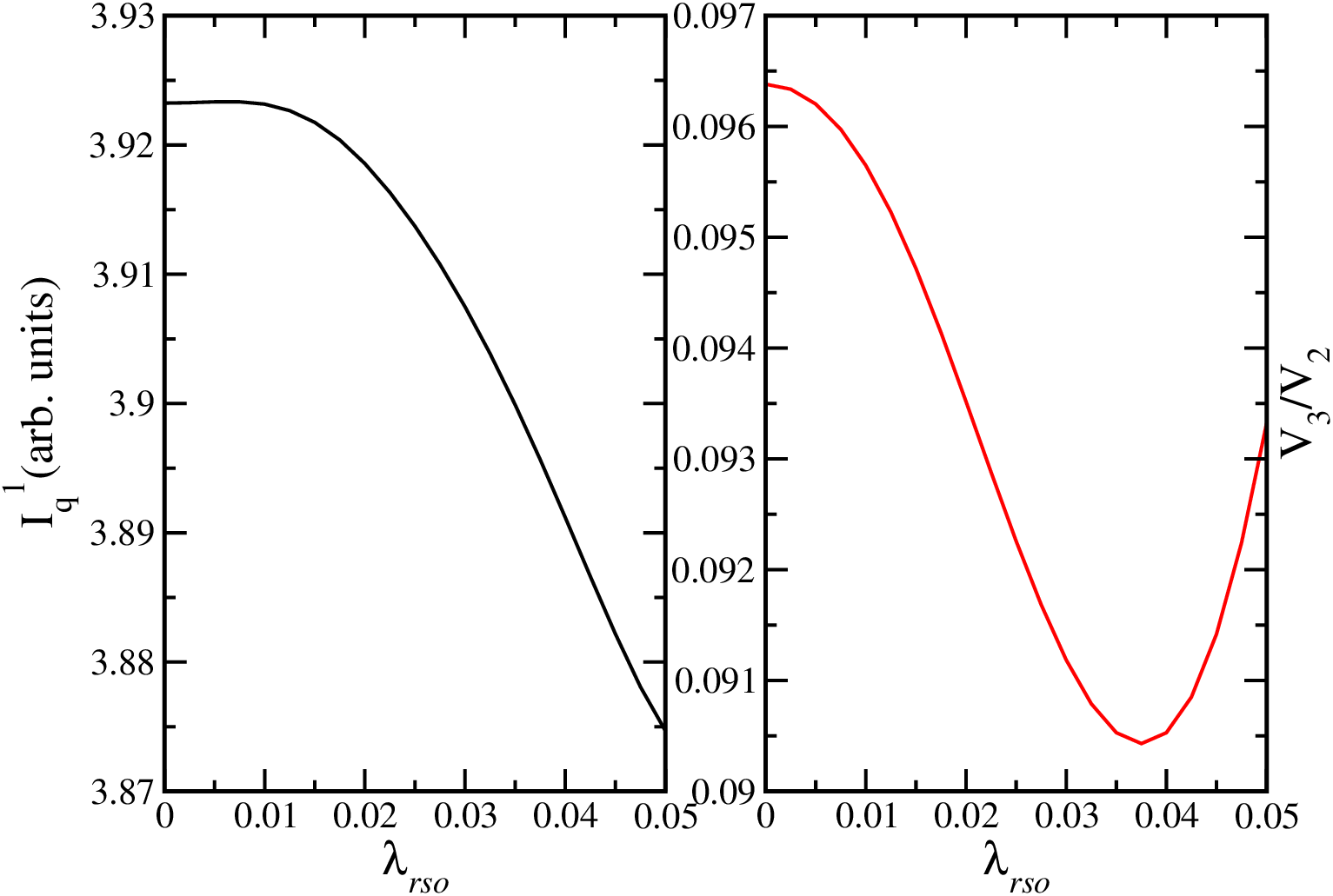}}
\caption{Charge current and voltage versus spin-orbit interaction strength 
$\lambda_{rso}$ for $E_{f}$=1.0t and $U$=2.0t.
 }
\label{Fig.2}
\end{figure}

%\vspace{0.5cm}
%\begin{figure}
%\resizebox{!}{2.in}{\includegraphics{neq_psc_NM_W0.eps}}
%\caption{spin current and voltage versus spin-orbit interaction strength 
%$\lambda_{rso}$ for 
%ballistic system. The other parameters are $E_{f}$=1.0t and $W$=0t. 
% }
%\label{Fig.2}
%\end{figure}

In Fig.2 we show charge current($I_{q}^{1}=-I_{q}^2$ 
flowing between terminal one and two and 
voltage at the
third terminal (which is voltage probe, i.e.,$I_{q}^{3}=0$ ) for ballistic system. We see that varies as spin orbit coupling is varied. Since the system is ballistic , hence the scattering occurs only due to spin-orbit interaction and accordingly the $V_{3}/V_{2}$ and $I_{q}^{1}$ shows variation. We notice that both charge current and voltage depends quadratically on the spin-orbit coupling strength $\lambda_{rso}$. This is expected since only non zero scattering matrix
element due to RSO are of second order in perturbation theory. Hence we see that in a three terminal geometry one can extract Rashba parameter just by measuring charge current or equivalently voltage. We stress that in two terminal
though the conductance will show variation as a function of Rashba parameter
, but these variation will be overshadowed due to beating pattern arising out of band mismatch at the contact between reservoirs and the system, which give rise to potential well at the contacts. Due to multiple reflections in this
potential well the conductance will show beating pattern arising out of interference between two Rashba spilt spin sub-bands \cite{pareek4}. Hence a two terminal measurement will suffer from the same difficulties and ambiguities as was the case for
SdH oscillation. Therefor a three terminal measurement is necessary and 
useful. 
\vspace{1cm}
\begin{figure}
\resizebox{!}{2.in}{\includegraphics{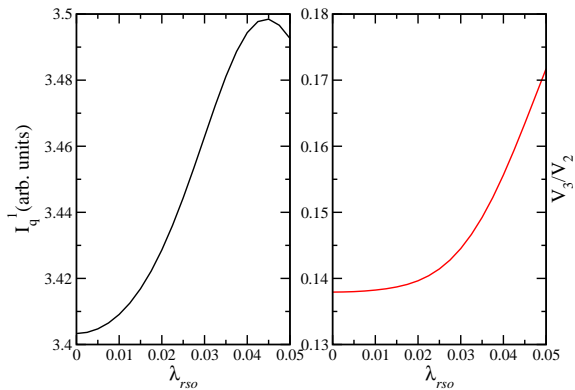}}
\caption{Charge current and voltage versus spin-orbit interaction strength 
$\lambda_{rso}$ for $E_{f}$=1.0t and $U$=0.5t.
 }
\label{Fig.3}
\end{figure}

In Fig.~3, Fig.~4 and Fig.~5 we present results for diffusive regime
corresponding to Anderson disorder strength U=0.5t, 1t and 2t , respectively.
For a given sample since disorder strength is fixed hence variations in charge current and voltages are just due to change in RSO strength as is seen in the Figures. However foe large values of $\lambda_{rso}$ there are qualitative changes between voltages and currents. This is so since we have not performed
disorder average, hence for a given distribution asymmetry is introduced between
different branches of Y-shaped conductor, which reflects in the qualitative difference between voltages and charge current. The asymmetry introduced for a
given distribution of impurities can be reduced by performing disorder averages
and hence the RSO strength can be calibrated for diffusive case as well.

\vspace{1cm}
\begin{figure}
\resizebox{!}{2.in}{\includegraphics{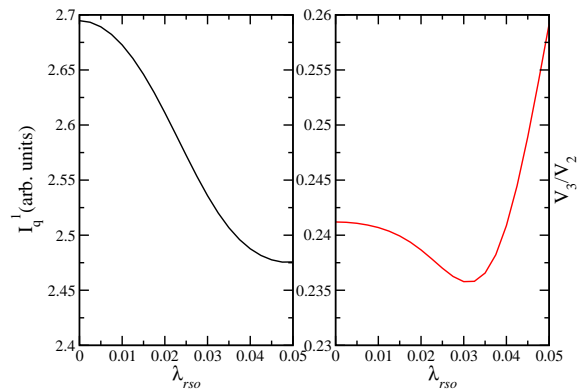}}
\caption{Charge current and voltage versus spin-orbit interaction strength 
$\lambda_{rso}$ for $E_{f}$=1.0t and $U$=1.0t.
 }
\label{Fig.4}
\end{figure}

\vspace{1cm}
\begin{figure}
\resizebox{!}{2.in}{\includegraphics{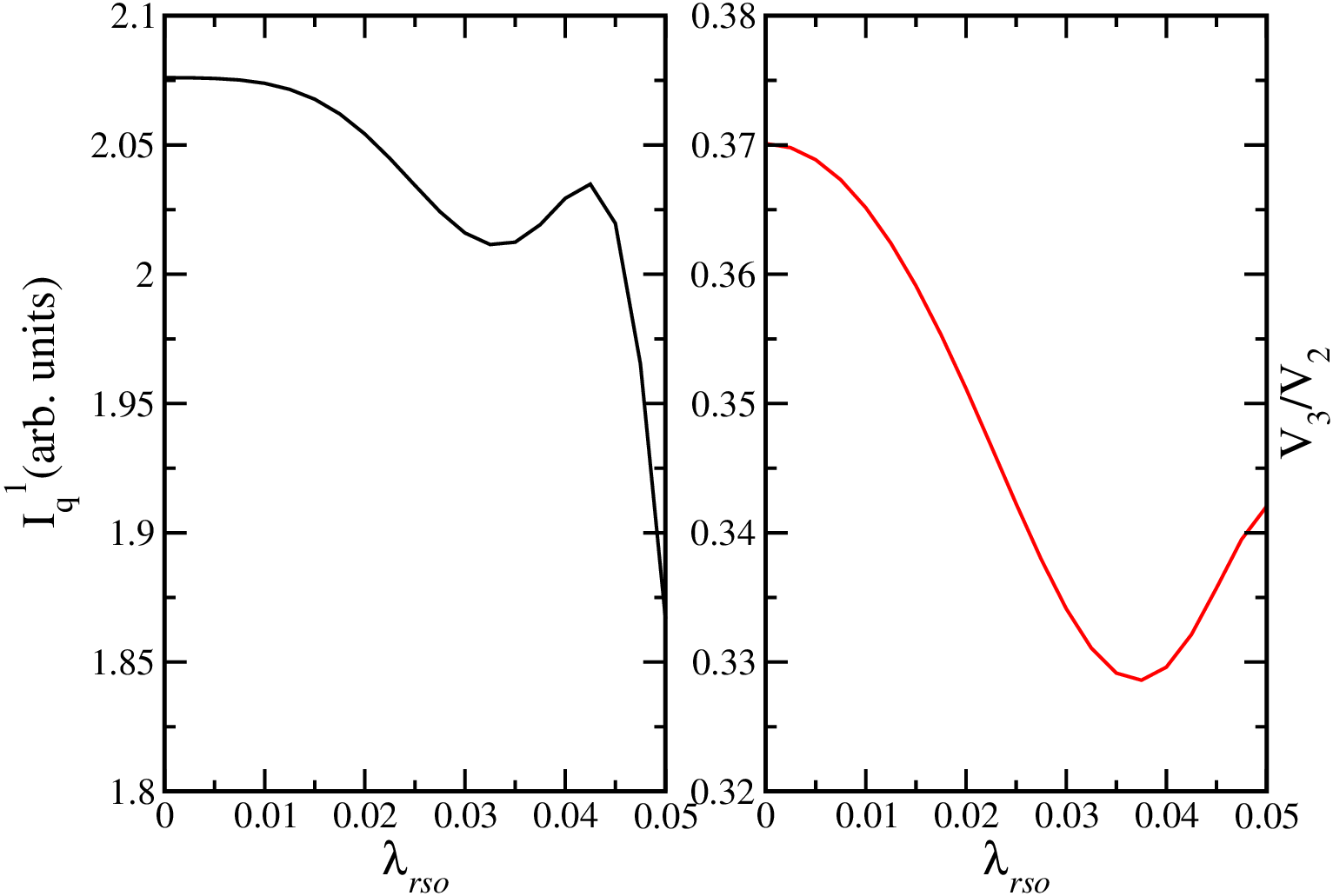}}
\caption{Charge current and voltage versus spin-orbit interaction strength 
$\lambda_{rso}$ for $E_{f}$=1.0t and $U$=2.0t.
 }
\label{Fig.5}
\end{figure}

%\vspace{1cm}
%\begin{figure}
%\resizebox{!}{2.in}{\includegraphics{neq_psc_NM_W.5.eps}}
%\caption{spin current and voltage versus spin-orbit interaction strength 
%$\lambda_{rso}$  for 
% $E_{f}$=1.0t and $U$=0.5t.
% }
%\label{Fig.3}
%\end{figure}

%\vspace{1cm}
%\begin{figure}
%\resizebox{!}{2.in}{\includegraphics{neq_psc_NM_W1.eps}}
%\caption{ spin current and voltage versus spin-orbit interaction strength 
%$\lambda_{rso}$ for $E_{f}$=1.0t and $U$=1.0t.
% }
%\label{Fig.4}
%\end{figure}

%\vspace{1cm}
%\begin{figure}
%\resizebox{!}{2.in}{\includegraphics{neq__psc_NM_W2.eps}}
%\caption{spin current and voltage versus spin-orbit interaction strength 
%$\lambda_{rso}$ for $E_{f}$=1.0t and $U$=2.0t.
% }
%\label{Fig.5}
%\end{figure}

Finally we discuss about the observebility of this effect. We see that for experimentally accessible regime of $\lambda_{rso}$ the current and voltage changes
respectively by 60$\%$ and 10$\%$. These changes are large enough to be observable in experiments. Further since changes in current $I_{q}^1$ is an order of magnitude larger compared to the changes in voltage, hence through charge measurement it will be easier to observe this effect and correspondingly measure Rashba
spin orbit interaction strength.

In conclusion we have proposed and shown that Rashba spin orbit interaction
can be measured in absence of magnetic field just by measuring charge current or voltages in a three terminal mesoscopic device. We have discussed advantages of this method over the standard SdH oscillation measurement for which applied
magnetic field is essential. Further the effect for charge current is of the order of 60$\%$ and for voltage it is of the order of 10$\%$. Hence by just measuring charge current in a three terminal mesoscopic system one can extract
Rashba spin orbit interaction strength. We hope that this will open up new opportunities in the field of spintronics.

\end{document}